\documentclass{PoS}

\title{IR subtraction schemes: integrating
the counterterms at NNLO in QCD}

\ShortTitle{IR subtraction schemes: integrating the counterterms at NNLO in QCD}

\author{\speaker{Paolo Bolzoni}
        \\
        DESY\\
        Platanenalle 6, D-15738 Zeuthen, Germany\\
        E-mail: \email{paolo.bolzoni@desy.de}}

\author{G\'abor Somogyi\\
        DESY\\
        Platanenalle 6, D-15738 Zeuthen, Germany\\
        E-mail: \email{gabor.somogyi@desy.de}}

\abstract{We briefly review a subtraction scheme
for computing radiative corrections to QCD jet cross sections
that can be defined at any
order in perturbation theory. Hereafter
we discuss the computational methods used to evaluate
analytically and numerically the integrated counterterms
arising from such a subtraction scheme. Basically these methods
the Mellin-Barnes (MB) representations technique together with the harmonic
summation and the sector
decomposition.}

\FullConference{13th International Workshop on Advanced Computing and Analysis Techniques in Physics Research, ACAT2010\\
		February 22-27, 2010\\
		Jaipur, India}

\begin{document}

\section{Introduction}

In quantum chromodynamics (QCD) and more generally in any
quantum field theory with massless particles one has to face among others
the problem of infrared (IR) divergences when computing higher
orders corrections. According to the KLN
(Kinoshita-Lee-Nauenberg) theorem
\cite{Kinoshita:1962ur,Lee:1964is}
these IR singularities cancel out once one put together all
the contributions at the same order which are degenerate
to a fixed final state (e.g. a parton state emitting
one collinear and/or soft gluon is degenerate
to the state without this extra emission). This means that
one has to compute the sum of several contributions which
are separately IR divergent leading to a final physical
and finite answer. To handle these IR singularities in a
general way is non-trivial already at the next-to-leading order
 (NLO) accuracy, where
however several solutions are known
\cite{Catani:1996vz,Frixione:1995ms,Giele:1991vf,Giele:1993dj,Nagy:1996bz,Frixione:1997np}.
In recent years a lot of effort has been devoted to
the extension to the NNLO accuracy
\cite{Weinzierl:2003fx,Frixione:2004is,GehrmannDeRidder:2005cm,Somogyi:2006da,Somogyi:2006db}.
In particular the subtraction scheme for QCD jet cross
sections defined in
\cite{Somogyi:2006da,Somogyi:2006db} is the one we are interested
in here. This scheme, initially
defined for processes without colored particles in the initial
state, has been extended to an NNLO-compatible scheme with hadronic
initial states \cite{Somogyi:2009ri}. Here we will consider only the
case of processes initiated by colorless particles and
we briefly summarize it as follows.

In QCD the perturbative expansion for any production rate at NNLO can
formally be written as
\begin{equation}
\sigma = \sigma^{\rm{LO}}+\sigma^{\rm{NLO}}+\sigma^{\rm{NNLO}}+\dots .
\end{equation}
Considering for example the $e^+ e^- \rightarrow m$ jet process we have
that the NNLO correction can be written as
\begin{equation}
\sigma^{\rm{NNLO}}=\int_{m+2}\,d\sigma^{\rm{RR}}_{m+2}J_{m+2}+
\int_{m+1}\,d\sigma^{\rm{RV}}_{m+1}J_{m+1}+
\int_{m}\,d\sigma^{\rm{VV}}_{m}J_{m}\,,
\label{nnlocorr}
\end{equation}
where the terms represent the doubly-real, the real-virtual and the
doubly-virtual contribution respectively. Each of them is IR divergent
while their sum remains finite. The restriction of the phase space
to define the physical quantity is realized by the jet functions
$J_n$.
The basic steps of subtraction
consist in  regularizing all the integrals in Eq.(\ref{nnlocorr})
using dimensional regularization in $d=4-2\epsilon$; then in reshuffling
the singularities between the three terms by adding and subtracting
suitable counterterms in such a way that we end up with three
contributions without IR singularities i.e. finite in $d=4$ dimensions.
In this way Eq.(\ref{nnlocorr}) becomes
\begin{equation}
\sigma^{\rm{NNLO}}=\int_{m+2}\,d\sigma^{\rm{NNLO}}_{m+2}+
\int_{m+1}\,d\sigma^{\rm{NNLO}}_{m+1}+
\int_{m}\,d\sigma^{\rm{NNLO}}_{m}.
\label{nnlocorrsub}
\end{equation}
Here
\begin{eqnarray}
d\sigma^{\rm{NNLO}}_{m+2}&=&
\left\{d\sigma^{\rm{RR}}_{m+2}J_{m+2}
-d\sigma^{\rm{RR,A_2}}_{m+2}J_{m}-
\left[
d\sigma^{\rm{RR,A_1}}_{m+2}J_{m+1}-d\sigma^{\rm{RR,A_{12}}}_{m+2}J_{m}
\right]
\right\}
\label{rr}\\
d\sigma^{\rm{NNLO}}_{m+1}&=&
\left[
d\sigma^{\rm{RV}}_{m+1}+\int_1\,d\sigma^{\rm{RR,A_1}}_{m+2}
\right]J_{m+1}-
\left[
d\sigma^{\rm{RV,A_1}}_{m+1}+\left(\int_1\,d\sigma^{\rm{RR,A_1}}_{m+2}\right)^{\rm{A_1}}
\right]J_{m}
\label{rv}\end{eqnarray}
and
\begin{equation}
d\sigma^{\rm{NNLO}}_{m}=\left\{
d\sigma^{\rm{VV}}_{m}+
\int_2\left[
d\sigma^{\rm{RR,A_2}}_{m+2}-d\sigma^{\rm{RR,A_{12}}}_{m+2}
\right]+
\int_1 \left[
d\sigma^{\rm{RV,A_1}}_{m+1}+\left(\int_1\,d\sigma^{\rm{RR,A_1}}_{m+2}\right)^{\rm{A_1}}
\right]\right\}J_{m}.
\label{vv}\end{equation}
In Eq.(\ref{rr}) above $d\sigma^{\rm{RR,A_1}}_{m+2}$ and
$d\sigma^{\rm{RR,A_2}}_{m+2}$ regularize the singly- and doubly-unresolved
limits of $d\sigma^{\rm{RR}}_{m+2}$ respectively. The last counterterm
in Eq.(\ref{rr}) which is $d\sigma^{\rm{RR,A_{12}}}_{m+2}$ must regularize
the sinlgy-unresolved limits of $d\sigma^{\rm{RR,A_2}}_{m+2}$ and the
doubly-unresolved limits of $d\sigma^{\rm{RR,A_1}}_{m+2}$.
Finally in Eq.(\ref{rv}) we have that
the counterterms $d\sigma^{\rm{RV,A_1}}_{m+2}$ and
$\left(\int_1\,d\sigma^{\rm{RR,A_1}}_{m+2}\right)^{\rm{A_1}}$
regularize the singly-unresolved limits of $d\sigma^{\rm{RV}}_{m+1}$
and $\int_1\,d\sigma^{\rm{RR,A_1}}_{m+2}$ respectively.

\section{Integrating the subtraction counterterms}

In order to complete the definition of the subtraction scheme,
one has to compute the integral of the counterterms
over the factorized one-
and two-body phase spaces given in Eqs.(\ref{rv},\ref{vv}).
These integrated counterterms have to be computed in $d=4-2\epsilon$ dimensions
and the result should be given in the form of a Laurent
expansion in $\epsilon$. According to the KLN theorem, the
$\epsilon$ poles of the $\epsilon$ expansions of the integrated
counterterms have to cancel those in the one-loop
correction $d\sigma^{\rm{RV}}_{m}$ and the two-loop correction
$d\sigma^{\rm{VV}}_{m}$ giving a finite contribution for
the real-virtual and doubly-real cross sections.
In this proceedings we will discuss the various techniques
used to compute the Laurent expansions in $\epsilon$ for the
integrated counterterms.

According to the different unresolved limits there are
different mappings of the external momenta all of them \emph{preserve
momentum conservation}.
The integrated singly-unresolved counterterm $\int_1\,d\sigma^{\rm{RR,A_1}}_{m+2}$
in Eq.(\ref{rv}) (
which is also the only one that is needed for a NLO computation) has been
computed in \cite{Somogyi:2006cz}.
As it is shown in \cite {Aglietti:2008fe,Bolzoni:2009ye}
the singly-unresolved integrated counterterm
$\int_1 [d\sigma^{\rm{RV,A_1}}_{m+1}+\left(\int_1\,d\sigma^{\rm{RR,A_1}}_{m+2}\right)^{\rm{A_1}}
]$ in Eq.(\ref{vv}) can be reduced to three different types of basic
integrals: these are the collinear, the soft and
the soft-collinear integrals .
Then the complete counterterm is built with them and their non
trivial convolutions (see e.g. Sections 2.1 and 2.2 of Ref.\cite{Bolzoni:2009ye}).
As far as the doubly-unresolved counterterm $\int_2[
d\sigma^{\rm{RR,A_2}}_{m+2}-d\sigma^{\rm{RR,A_{12}}}_{m+2}]$ in Eq.(\ref{vv})
is concerned we have that the computation of the second one
will be published in a forthcoming paper \cite{bolzoni:preparation}.
The computation of the first term is clearly feasible employing
the same methods and is a work in progress.
For $\int_2d\sigma^{\rm{RR,A_{12}}}_{m+2}$  it turned out that it can be
reduced to six different basic integrals divided in couples of
collinear, soft and collinear-soft integrals. Analogously to the singly-unresolved
case this integrated counterterm it is built up with these
basic integrals and their non-trivial
convolutions.

To compute the basic integrals that are involved in the definitions of
the various integrated counterterms we studied different methods.
A first fully numerical method for the computation of the basic
integrals is based on sector decompisition (see \cite{Heinrich:2008si}
and references therein).
To extract the poles in $\epsilon$ of these integrals
a \texttt{Mathematica} package has
been written implementing the sector decomposition techniques
\cite{Aglietti:2008fe,Somogyi:2008fc}. This program also produces
\texttt{FORTRAN} codes directly used by numerical integration programs.
Another method that has been used to approach the computation
of these integrals exploited integration-by-parts (IBP) identities.
With this method analytical results were obtained for some of the
singly-unresolved integrals \cite{Aglietti:2008fe}. However it turned
out that the method based on Mellin-Barnes (MB)
representations is in many cases more efficient and accurate
than the numerical evaluation via sector decomposition and
made the analytic results much more feasible than the techniques
based on IBP  \cite{Bolzoni:2009ye}. Both the sector decomposition and
the MB techniques have been implemented independently to cross
check all the results (also for those involved in
the doubly-unresolved counterterm) obtained for the Laurent expansions.

To briefly show how the MB method works we want to describe a representative
example involved in the computation of the  $\int_2d\sigma^{\rm{RR,A_{12}}}_{m+2}$
counterterm. As anticipated a full treatment of it will
appear elsewhere \cite{bolzoni:preparation}.
We begin by considering the following basic collinear integral
\begin{equation}
I_k(x)=x\, \int_0^1 \,d\alpha dv \frac{\alpha^{-1-\epsilon}(1-\alpha)^{2d_0-1}
v^{-\epsilon}(1-v)^{-\epsilon}}
{[\alpha+(1-\alpha)x]^{1+\epsilon}}
\left(\frac{\alpha+(1-\alpha)x v}{2\alpha+(1-\alpha)x}\right)^k,
\label{coll}
\end{equation}
where $d_0\geq 2$ is an arbitrary parameter, $k=-1,0,1,2$ and $x$ is a
kinematic variable in $[0,1]$. Let us now consider
the particular case $d_0=2, k=1$. With this choice the integral in
Eq.(\ref{coll}) is split into two different contributions. Of them only
the second one produces a pole in $\epsilon$ due to the factor
$\alpha^{-1-\epsilon}$ and we concentrate on it, which explicitly is given
by
\begin{equation}
E(x)=x^2\, \int_0^1\,d\alpha
\frac{\alpha^{-1-\epsilon}(1-\alpha)^{4}
}
{[\alpha+(1-\alpha)x]^{1+\epsilon}[2\alpha+(1-\alpha)x]}\,,
\label{example}
\end{equation}
where we have neglected an overall factor of
$\Gamma(2-\epsilon)\Gamma(1-\epsilon)/\Gamma(3-2\epsilon)$
coming from the integration over $v$.
Now applying the well known basic MB formula
\begin{equation}
\frac{1}{(a+b)^\nu}=\frac{1}{\Gamma(\nu)}\int_{-\rm i\infty}^{+\rm i\infty}
\frac{dz}{2\pi i}\,a^{-\nu-z}b^{z}\,\Gamma(\nu+z)\Gamma(-z)
\label{mbbasic}
\end{equation}
two times in Eq.(\ref{example}) we obtain
\begin{equation}
E(x)=\int_{-\rm i\infty}^{+\rm i\infty}\frac{dz_1 dz_2}{(2\pi i)^2}
\,2^{z_2}x^{-\epsilon-z_1-z_2}
\frac{\Gamma(-z_1)\Gamma(-z_2)\Gamma(3-\epsilon-z)
\Gamma(1+\epsilon+z_1)\Gamma(1+z_2)\Gamma(-\epsilon+z)}
{\Gamma(3-2\epsilon)\Gamma(1+\epsilon)},
\label{mbex}
\end{equation}
where $z=z_1+z_2$. According to the definition of the complex integration
path in Eq.(\ref{mbbasic}), one has to choose the contours in such a way
that the poles with a $\Gamma(\dots +z)$ dependence are to the left of the
contour and the poles with a $\Gamma(\dots -z)$ dependence are to the
right of it. Clearly if one starts with contours that do not satisfy
this requirement one has to deform them taking into account the
residua of the integrand while crossing every pole of it.
This procedure is automatized in the \texttt{Mathematica} package
\texttt{MB.m} \cite{Czakon:2005rk}. Once this is done this package
enables one to easily compute the Laurent expansion in $\epsilon$
where the coefficients are given as a list of MB integrals.
The numerical evaluation of the integrals is also implemented in
\texttt{MB.m} by use of a simple command line. Moreover according to
the Cauchy theorem the integration over the complex contour can
be converted into sums over the residua inside the paths. If we
follow these steps for our example given in Eq.(\ref{mbex}) we get
\begin{eqnarray}
E(x)&=&-\frac{1}{\epsilon}+2\log\left(\frac{x}{2}\right)-
\log(2)\sum_{n=1}^\infty x^n {n+2 \choose 2}-
\sum_{m,n=1}^\infty \left(\frac{x}{2}\right)^m x^n {m+n+2 \choose 2}\times \nonumber\\
&&\times \left[\rm S_1(m+n+2)-\rm S_1(m+n)+\log\left(\frac{x}{2}\right)\right]
+\emph{O}(\epsilon),
\label{sumex}
\end{eqnarray}
where $\rm S_1(n)=\sum_{i=1}^n \frac{1}{i}$ denotes the usual harmonic numbers. To obtain
an analytic answer the harmonic summation involved in Eq.(\ref{sumex}) has to
be performed. In this example, like in many other cases encountered in the
complete computation of the integrated counterterms, the harmonic summation
is feasible and can be performed using the \texttt{XSummer} package
\cite{Moch:2001zr,Moch:2005uc} in \texttt{Form 3.0} \cite{Vermaseren:2000nd}.
Running a proper script we obtain for our example in Eq.(\ref{sumex}) the following
analytic answer:
\begin{eqnarray}
E(x)&=&-\frac{1}{\epsilon}+\log(2)\left(1-\frac{1}{(1-x)^3}\right)-
\frac{x^2 (3x^2-15x+14)}{2(1-x)^2(2-x)^2}\nonumber\\
&&+\frac{(x^6-9x^5+33x^4-78x^3+108x^2-72x+16)}{(1-x)^3(2-x)^3}+
\emph{O}(\epsilon).
\end{eqnarray}
Note that the limit in $x=1$ (corresponding to the collinear limit)
is well defined because we have that $\lim_{x\to 1}E=-1/\epsilon
+53/6-16\log 2 +\emph{O}(\epsilon)$. This concludes the discussion of our
example to show the mehtod of MB representations.

As mentioned above, in the forthcoming paper about the counterterm
$\int_2d\sigma^{\rm{RR,A_{12}}}_{m+2}$ we will also involve
convolutions of the basic integrals, for example the collinear one
in Eq.(\ref{coll}) with itself:
\begin{equation}
I_k * I_l(x,y)=y\, \int_0^1 \,d\alpha dv \frac{\alpha^{-1-\epsilon}
(1-\alpha)^{2d_0-3+2\epsilon}
v^{-\epsilon}(1-v)^{-\epsilon}}
{[\alpha+(1-\alpha)y]^{1+\epsilon}}
\left(\frac{\alpha+(1-\alpha)y v}{2\alpha+(1-\alpha)y}\right)^l
I_k ((1-\alpha)x).
\end{equation}
For this particular integral following the steps described above we find
\begin{eqnarray}
I_k * I_l(x,y)&=&\frac{\delta_{k,-1}\delta_{l,-1}}{4\epsilon^4}
     -\left[\frac{\delta_{k,-1}\delta_{l,-1}}{2} \ln (x y)
     +\frac{(1-\delta_{k,-1})\delta_{l,-1}}{2(1+k(1-\delta_{k,-1}))}
     +\frac{(1-\delta_{l,-1})\delta_{k,-1}}{2(1+l(1-\delta_{l,-1}))}\right]
     \frac{1}{\epsilon^3}\nonumber\\
     &&     + \emph{O}(\epsilon^{-2}).
\end{eqnarray}
Here the $\emph{O}(\epsilon^{-2})$ contribution is already cumbersome
and will not be given here.

\section{Conclusions}

In this proceedings after shortly reviewing the NNLO subtraction scheme developed
in \cite{Somogyi:2006da,Somogyi:2006db,Somogyi:2009ri} we have discussed the
various methods to compute the integrated counterterms focussing on the
MB representations method.
Showing  a simple example we described its usage discussing some new
results: the computation of the integrated counterterm
$\int_2d\sigma^{\rm{RR,A_{12}}}_{m+2}$ is finished and the results
will be reported in \cite{bolzoni:preparation}. Our methods are clearly
applicable to compute the final remaining integrated counterterm
$\int_2d\sigma^{\rm{RR,A_{2}}}_{m+2}$ which is a work in progress.
We found that the method based on
the MB representations can be used to obtain analytic results
and in many cases is more efficient even numerically than others like
sector decomposition.


\begin{thebibliography}{10}

\bibitem{Kinoshita:1962ur}
T.~Kinoshita.
\newblock {Mass singularities of Feynman amplitudes}.
\newblock {\em J. Math. Phys.}, 3:650--677, 1962.

\bibitem{Lee:1964is}
T.~D. Lee and M.~Nauenberg.
\newblock {Degenerate Systems and Mass Singularities}.
\newblock {\em Phys. Rev.}, 133:B1549--B1562, 1964.

\bibitem{Catani:1996vz}
S.~Catani and M.~H. Seymour.
\newblock {A general algorithm for calculating jet cross sections in NLO QCD}.
\newblock {\em Nucl. Phys.}, B485:291--419, 1997, hep-ph/9605323.

\bibitem{Frixione:1995ms}
S.~Frixione, Z.~Kunszt, and A.~Signer.
\newblock {Three jet cross-sections to next-to-leading order}.
\newblock {\em Nucl. Phys.}, B467:399--442, 1996, hep-ph/9512328.

\bibitem{Giele:1991vf}
W.~T. Giele and E.~W.~Nigel Glover.
\newblock {Higher order corrections to jet cross-sections in e+ e-
  annihilation}.
\newblock {\em Phys. Rev.}, D46:1980--2010, 1992.

\bibitem{Giele:1993dj}
W.~T. Giele, E.~W.~Nigel Glover, and David~A. Kosower.
\newblock {Higher order corrections to jet cross-sections in hadron colliders}.
\newblock {\em Nucl. Phys.}, B403:633--670, 1993, hep-ph/9302225.

\bibitem{Nagy:1996bz}
Zoltan Nagy and Zoltan Trocsanyi.
\newblock {Calculation of QCD jet cross sections at next-to-leading order}.
\newblock {\em Nucl. Phys.}, B486:189--226, 1997, hep-ph/9610498.

\bibitem{Frixione:1997np}
S.~Frixione.
\newblock {A General approach to jet cross-sections in QCD}.
\newblock {\em Nucl. Phys.}, B507:295--314, 1997, hep-ph/9706545.

\bibitem{Weinzierl:2003fx}
Stefan Weinzierl.
\newblock {Subtraction terms at NNLO}.
\newblock {\em JHEP}, 03:062, 2003, hep-ph/0302180.

\bibitem{Frixione:2004is}
Stefano Frixione and Massimiliano Grazzini.
\newblock {Subtraction at NNLO}.
\newblock {\em JHEP}, 06:010, 2005, hep-ph/0411399.

\bibitem{GehrmannDeRidder:2005cm}
A.~Gehrmann-De~Ridder, T.~Gehrmann, and E.~W.~Nigel Glover.
\newblock {Antenna Subtraction at NNLO}.
\newblock {\em JHEP}, 09:056, 2005, hep-ph/0505111.

\bibitem{Somogyi:2006da}
Gabor Somogyi, Zoltan Trocsanyi, and Vittorio Del~Duca.
\newblock {A subtraction scheme for computing QCD jet cross sections at NNLO:
  regularization of doubly-real emissions}.
\newblock {\em JHEP}, 01:070, 2007, hep-ph/0609042.

\bibitem{Somogyi:2006db}
Gabor Somogyi and Zoltan Trocsanyi.
\newblock {A subtraction scheme for computing QCD jet cross sections at NNLO:
  regularization of real-virtual emission}.
\newblock {\em JHEP}, 01:052, 2007, hep-ph/0609043.

\bibitem{Somogyi:2009ri}
Gabor Somogyi.
\newblock {Subtraction with hadronic initial states: an NNLO- compatible
  scheme}.
\newblock {\em JHEP}, 05:016, 2009, 0903.1218.

\bibitem{Somogyi:2006cz}
Gabor Somogyi and Zoltan Trocsanyi.
\newblock {A new subtraction scheme for computing QCD jet cross sections at
  next-to-leading order accuracy}.
\newblock 2006, hep-ph/0609041.

\bibitem{Aglietti:2008fe}
Ugo Aglietti, Vittorio Del~Duca, Claude Duhr, Gabor Somogyi, and Zoltan
  Trocsanyi.
\newblock {Analytic integration of real-virtual counterterms in NNLO jet cross
  sections I}.
\newblock {\em JHEP}, 09:107, 2008, 0807.0514.

\bibitem{Bolzoni:2009ye}
Paolo Bolzoni, Sven-Olaf Moch, Gabor Somogyi, and Zoltan Trocsanyi.
\newblock {Analytic integration of real-virtual counterterms in NNLO jet cross
  sections II}.
\newblock {\em JHEP}, 08:079, 2009, 0905.4390.

\bibitem{bolzoni:preparation}
Paolo Bolzoni, Gabor Somogyi, and Zoltan Trocsanyi.
\newblock {(in preparation)}.

\bibitem{Heinrich:2008si}
Gudrun Heinrich.
\newblock {Sector Decomposition}.
\newblock {\em Int. J. Mod. Phys.}, A23:1457--1486, 2008, 0803.4177.

\bibitem{Somogyi:2008fc}
Gabor Somogyi and Zoltan Trocsanyi.
\newblock {A subtraction scheme for computing QCD jet cross sections at NNLO:
  integrating the subtraction terms I}.
\newblock {\em JHEP}, 08:042, 2008, 0807.0509.

\bibitem{Czakon:2005rk}
M.~Czakon.
\newblock {Automatized analytic continuation of Mellin-Barnes integrals}.
\newblock {\em Comput. Phys. Commun.}, 175:559--571, 2006, hep-ph/0511200.

\bibitem{Moch:2001zr}
Sven Moch, Peter Uwer, and Stefan Weinzierl.
\newblock {Nested sums, expansion of transcendental functions and multi-scale
  multi-loop integrals}.
\newblock {\em J. Math. Phys.}, 43:3363--3386, 2002, hep-ph/0110083.

\bibitem{Moch:2005uc}
S.~Moch and P.~Uwer.
\newblock {XSummer: Transcendental functions and symbolic summation in Form}.
\newblock {\em Comput. Phys. Commun.}, 174:759--770, 2006, math-ph/0508008.

\bibitem{Vermaseren:2000nd}
J.~A.~M. Vermaseren.
\newblock {New features of FORM}.
\newblock 2000, math-ph/0010025.

\end{thebibliography}

\end{document}